\documentclass[prl,aps,twocolumn,showpacs, tightinlines]{revtex4}%
\usepackage{amsmath}
\usepackage{amsfonts}
\usepackage{amssymb}
\usepackage{graphicx}
\usepackage{bm}

\newcommand{\etl}{\textit{et al.}}

\begin{document}
\title{Non-exponential London penetration depth in RFeAsO$_{0.9}$F$_{0.1}$ (R=La,Nd) single crystals}

\author{C.~Martin}
\affiliation{Ames Laboratory and Department of Physics \& Astronomy, Iowa State University,
Ames, IA 50011}
\author{M.~E.~Tillman}
\affiliation{Ames Laboratory and Department of Physics \& Astronomy, Iowa State University,
Ames, IA 50011}
\author{H.~Kim}
\affiliation{Ames Laboratory and Department of Physics \& Astronomy, Iowa State University,
Ames, IA 50011}
\author{M.~A.~Tanatar}
\affiliation{Ames Laboratory and Department of Physics \& Astronomy, Iowa State University,
Ames, IA 50011}
\author{S.~K.~Kim}
\affiliation{Ames Laboratory and Department of Physics \& Astronomy, Iowa State University,
Ames, IA 50011}
\author{A.~Kreyssig}
\affiliation{Ames Laboratory and Department of Physics \& Astronomy, Iowa State University,
Ames, IA 50011}
\author{R.~T.~Gordon}
\affiliation{Ames Laboratory and Department of Physics \& Astronomy, Iowa State University,
Ames, IA 50011}
\author{M.~D.~Vannette}
\affiliation{Ames Laboratory and Department of Physics \& Astronomy, Iowa State University,
Ames, IA 50011}
\author{S.~Nandi}
\affiliation{Ames Laboratory and Department of Physics \& Astronomy, Iowa State University,
Ames, IA 50011}
\author{V.~G.~Kogan}
\affiliation{Ames Laboratory and Department of Physics \& Astronomy, Iowa State University,
Ames, IA 50011}
\author{S.~L.~Bud'ko}
\affiliation{Ames Laboratory and Department of Physics \& Astronomy, Iowa State University,
Ames, IA 50011}
\author{P.~C.~Canfield}
\affiliation{Ames Laboratory and Department of Physics \& Astronomy, Iowa State University,
Ames, IA 50011}
\author{A.~I.~Goldman}
\affiliation{Ames Laboratory and Department of Physics \& Astronomy, Iowa State University,
Ames, IA 50011}
\author{R.~Prozorov}
\email[Corresponding author: ]{prozorov@ameslab.gov}
\affiliation{Ames Laboratory and Department of Physics \& Astronomy, Iowa State University,
Ames, IA 50011}

\date{12 March 2009}

\begin{abstract}
The superconducting penetration depth, $\lambda(T)$, has been measured in RFeAsO$_{0.9}$F$_{0.1}$ (R=La,Nd) single crystals (R-1111). In Nd-1111, we find an upturn in $\lambda(T)$ upon cooling and attribute it to the paramagnetism of the Nd ions, similar to the case of the electron-doped cuprate Nd-Ce-Cu-O. After the correction for paramagnetism, the London penetration depth variation is found to follow a power-law behavior, $\Delta \lambda_L(T)\propto T^{2}$ at low temperatures. The same $T^2$ variation of $\lambda(T)$ was found in non-magnetic La-1111 crystals. Analysis of the superfluid density and of penetration depth anisotropy over the full temperature range is consistent with two-gap superconductivity. Based on this and on our previous work, we conclude that both the RFeAsO (1111) and BaFe$_2$As$_2$ (122) families of pnictide superconductors exhibit unconventional two-gap superconductivity.
\end{abstract}

\maketitle
A year after the discovery of iron pnictide superconductors \cite{Kamihara08}, the order parameter (OP) symmetry still remains one of the most important open questions. The pnictides are complex superconductors with a possible influence from magnetism, similar to the cuprates~\cite{Zhao08}, and possible multigap superconductivity, as observed in MgB$_2$~\cite{Mazin03}. Theoretical reports predict that a spin-density wave instability and multiple Fermi surface sheets can give rise to a complex momentum dependence of the superconducting OP. It can have opposite signs on different sheets of the Fermi surface (FS) and have \textit{s}-wave ($s^{\pm}$-wave)~\cite{Mazin08}, anisotropic $s$-wave~\cite{Rastko08}, extended \textit{s}-wave with nodes \cite{Graser08}, or $d_{x^{2}-y^{2}}$-wave symmetry ~\cite{Seo08}.

The London penetration depth, $\lambda_L(T)$, is among the most useful probes of the OP symmetry \cite{Prozorov2006}. A fully gapped FS leads to exponential saturation of $\lambda_L(T \rightarrow 0)$. The saturation temperature is determined by the minimum value of the gap, which can either be due to pairing anisotropy or/and different gap amplitudes on different sheets of the FS. The presence of the gap zeroes (point or line nodes) on the FS leads to a non-exponential variation of $\lambda_L(T)$, e.g. power-law behavior. In the high-$T_c$ cuprates, precision measurements of $\lambda_L(T)$ were the first to convincingly show the existence of \textit{d}-wave pairing state ~\cite{Hardy93}. Measurements of $\lambda(T)$ in recently discovered FeAs-compounds reveal a contradictory picture. Extensive studies by our group on both electron-doped  Ba(Fe,Co)$_2$As$_2$~\cite{Gordon08} and hole-doped (Ba,K)Fe$_2$As$_2$~\cite{Martin09} (122) systems have found $\lambda_L(T) \propto T^2$ behavior down to $\approx0.02T_{c}$, which may be a signature of unconventional $s^{\pm}$ state ~\cite{Parish08, Vorontsov09}. Microwave measurements in hole-doped 122 suggest a fully gapped state with two gaps \cite{Hashimoto08}, whereas $\mu$SR measurements show a linear temperature dependence of the superfluid density, suggesting a nodal gap~\cite{Goko08}. In R-1111, an exponential dependence in $\lambda(T)$ was reported in single crystals of Sm-1111 \cite{Malone08} and Pr-1111 \cite{Hashimoto09,okazaki09} as well as in our own experiments on Nd-1111 \cite{Martin08}. A nodeless gap in R-1111 is supported by ARPES~\cite{Kondo08} and by point contact spectroscopy with single ~\cite{Chen08} and two-gaps \cite{Samuely09}, whereas $\mu$SR~\cite{Luetkens350nm} and NQR \cite{Kawasaki08} results could be interpreted as both, multigap \textit{s}- or \textit{d}-wave state. Furthermore, penetration depth measurements in stoichiometric 1111 LaFePO suggest the existence of line nodes~\cite{Fletcher08}.

\begin{figure}[tb]
\includegraphics[width=8.5cm]{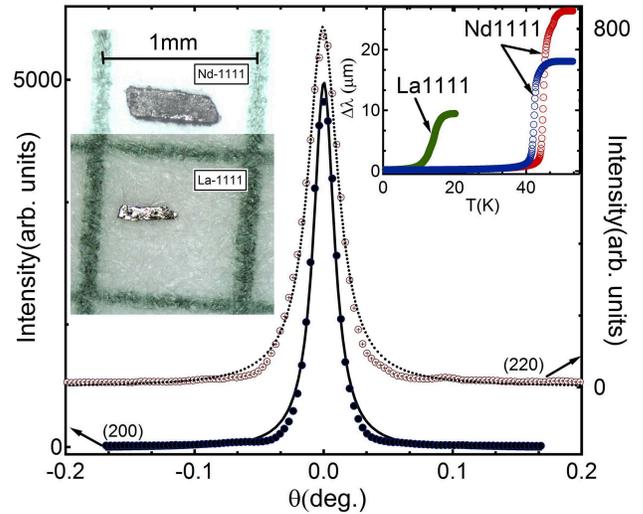}
\caption{(Color online) Transverse scans through the (200) and (220) Bragg reflections of the Nd-1111 single crystal. The lines are fits to Lorentzians yielding a full width half maximum of 0.019~deg and 0.028~deg, respectively. Left inset: photographs of Nd-1111  (top) and La-1111 (bottom) single crystals. Right inset: $\lambda(T)$ in the full temperature range for three samples.}
 \label{fig:Xray}
 \end{figure}

In this paper we report on penetration depth measurements in RFeAsO$_{0.9}$F$_{0.1}$ (R=La, Nd) single crystals. The Nd-1111 crystals were much larger than those used in our previous study~\cite{Martin08} and therefore our new data were obtained with notably improved signal-to-noise (S/N) ratio. We find that the measured $\lambda(T)$ is not exponentially flat, but shows a paramagnetic upturn caused by the Nd$^{3+}$ ions. Similar situation was resolved ten years ago in the electron doped cuprate Nd$_{1.85}$Ce$_{0.15}$CuO$_{4-x}$ (NCCO) \cite{Cooper96,Kokales00}. Correcting the data for this paramagnetic contribution, we have determined the London penetration depth, $\lambda_L(T)$, which does not saturate down to 0.02$T_c$ and is best described by the power-law $\Delta \lambda_{L}(T)\propto T^{2}$. Measurements on the recently grown non-magnetic La-1111 crystals reveal the same power-law without any corrections. The anisotropy of the penetration depth, $\gamma_{\lambda}=\lambda_{c}/\lambda_{ab}$, was found to increase upon cooling, similar to the 122 compounds \cite{Martin09,Prozorov09}, which combined with the positive curvature of the superfluid density at elevated temperatures may be considered a consequence of multiband superconductivity.

Single crystals of RFeAsO$_{0.9}$F$_{0.1}$ (nominal O and F composition) were extracted from 5 mm diameter pellets obtained using high pressure synthesis ~\cite{Prozorov08}. Individual superconducting single crystals have dimensions up to $650 \times 180 \times 120$ $\mu$m$^{3}$ for Nd-1111 and $330 \times 240 \times 10$ $\mu$m$^{3}$ for La-1111 (left inset of Fig.~\ref{fig:Xray}). In the following, we present results for the largest samples with the highest S/N ratio. Similar results were obtained in four other somewhat smaller crystals. The mean $T_c$ was 45 K for Nd-1111 and 14 K for La-1111 crystals.

The samples were characterized using synchrotron X-rays (6ID-D beamline in the MUCAT sector at the Advanced Photon Source, Argonne) with an energy of 99.6~keV and an absorption length of 1.5~mm, probing throughout the sample thickness. The incident beam was collimated to 0.1x0.1~mm$^2$ and the sample {\bf c}-direction was aligned parallel to the beam. Two-dimensional scattering patterns were measured with a MAR345 image-plate positioned 1705 mm behind the sample. Complete reciprocal planes were recorded by rocking the sample during the exposure through two angles perpendicular to the beam over a range of $\pm$~2.4~deg ~\cite{Kreyssig07}. This way the Bragg reflections within reciprocal planes perpendicular to the incoming beam were recorded in a single exposure of the detector (typically 3.5 min). Stepwise translations of the sample perpendicular to the incident beam allowed measurements of the spatially resolved diffraction patterns in the basal {\bf ab}-plane. The entire sample was scanned and no impurities, misoriented grains or diffuse signals from disordered material were detected. The excellent quality of the single crystals used in this study can be seen from transverse scans, at room temperature, through selected Bragg reflections extracted from the two-dimensional pattern, Fig.~\ref{fig:Xray}. We find sharp peaks with widths of 0.019~deg and 0.028~deg, typical of high quality single crystals. The diffraction pattern did not change qualitatively on cooling down to 15~K, revealing no traces of the structural transition that is observed in the non-superconducting parent  compound.

\begin{figure}[tb]
\begin{center}
\includegraphics[width=8.5cm]{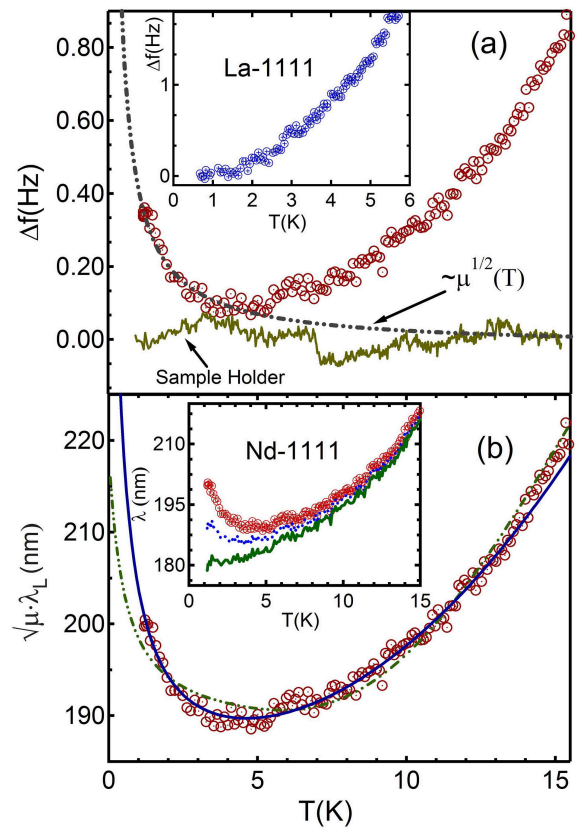}
\end{center}
\caption{(Color online) (a) Frequency shift $\Delta f(T)$ for a Nd-1111 crystal (symbols), contribution from the sample holder (continuous line) and $\sqrt{\mu}$ (dashed line). Inset: $\Delta f(T)$ for the La-1111 crystal. (b) Low-temperature region of $\lambda(T)$ in the Nd-1111 crystal. The lines are fits to Eq.~(\ref{Eq: BCS_Mu}) (dashed) and Eq.~(\ref{Eq: Pow_Mu}) (continuous) Inset: Original data (symbols) and $\lambda_{L}$(T) obtained after dividing by $\sqrt{\mu}$ from a fit to Eq.~(\ref{Eq: BCS_Mu}) (dots) and  to Eq.~(\ref{Eq: Pow_Mu}) (continuous line).}
\label{fig:DivCurr}
\end{figure}

The magnetic penetration depth, $\lambda(T)$, was measured by placing the sample inside a 1 $\mu$H inductor of a self-resonating tunnel-diode resonator (TDR) with resonant frequency $f_{0}=1/2\pi\sqrt{LC} \approx 14$ MHz. The excitation ac magnetic field, $H_{ac}\sim 10$ mOe, is much smaller than the lower critical field $H_{c1}\sim 100$ Oe, assuring that the sample is in the Meissner state. The measured quantity is the shift in the resonant frequency, $\Delta f \equiv f(T)-f_{0} =-4 \pi\chi(T)G$, where $\chi$ is the total magnetic susceptibility and $G\simeq f_{0}V_{s}/2V_{c}\left( 1-N\right) $ is a geometric calibration factor defined by the coil, $V_c$, and the sample,  $V_{s}$, volumes and the demagnetization factor $N$. $G$ is measured directly by pulling the sample out of the coil at the lowest temperature \cite{Prozorov2006}. The susceptibility in the Meissner state can be written in terms of $\lambda_L(T)$, $\mu(T)$, and a characteristic sample dimension $w$ as $4\pi\chi\left( T\right) =\left[\sqrt{\mu(T)} \lambda_L(T)/w\right]\,\tanh{\left[ \sqrt{\mu}w/\lambda_L(T)\right] }-1$, where $\mu(T)$ is the normal state paramagnetic permeability ~\cite{Prozorov2000,Prozorov2006}.

Figure~\ref{fig:DivCurr}(a) shows the frequency shifts for Nd-1111 (main frame) and La-1111 (inset) crystals at low temperatures (the whole range is shown in the inset in Fig.\ref{fig:Xray}). The samples were mounted with $H_{ac}\parallel c$, so that $\Delta f\propto \Delta\lambda_{ab}$. Note that the total frequency change over 15 K is about 1 Hz, which is less than 0.1 ppm. However, this is still significantly larger than the noise level of our system, shown in Fig.~\ref{fig:DivCurr}(a) for comparison. Whereas the data in La-1111 are monotonic, the measured $\Delta f(T)$ in Nd-1111 shows an upturn below 4K. A similar upturn in the electron-doped cuprate NCCO was explained by the local moment magnetism of the Nd ions ~\cite{Cooper96,Kokales00}. In the presence of paramagnetic ions, the screening length is reduced to $\lambda(T,\mu)=\lambda_L(T)/\sqrt{\mu(T)}$, where $\lambda_L$ is the London penetration depth. At low temperatures, where $\tanh(w/\lambda) \approx 1$, the measured $\Delta f$ can be written as $\Delta f(T)=G\sqrt{\mu}\lambda_{L}(T)$ \cite{Cooper96}. If we assume Currie-Weiss behavior for the magnetic susceptibility, $\chi(T)=C/(T+\theta)$, with $C$ being the Curie constant ($4\pi$ factor included) and $\theta$ being the Curie-Weiss temperature, we obtain an upturn in $\Delta f(T)$ provided the paramagnetic increase is larger than the decrease due to $\lambda_{L}(T)$. Figure~\ref{fig:DivCurr}(a) shows that $\sqrt{\mu(T)}$ clearly dominates the experimental data below 4 K. To subtract the contribution of the paramagnetic term, we fit the data assuming $\lambda_L(T)$ to be either exponential, Eq.~(\ref{Eq: BCS_Mu}), or power law, Eq.~(\ref{Eq: Pow_Mu}),

\begin{subequations}
\begin{align}
\lambda(T)=\sqrt{\mu(T)}\lambda(0)\left[1+\sqrt
{\frac{\pi\Delta_0  }{2T}}\exp\left(  -\frac{\Delta_0 }{T}\right)\right]\label{Eq: BCS_Mu}\\
\lambda(T)=\sqrt{\mu(T)}\lambda(0)\left[1+AT^{n}\right]\label{Eq: Pow_Mu}
\end{align}
\end{subequations}

\begin{figure}[tb]
\includegraphics[width=8.5cm]{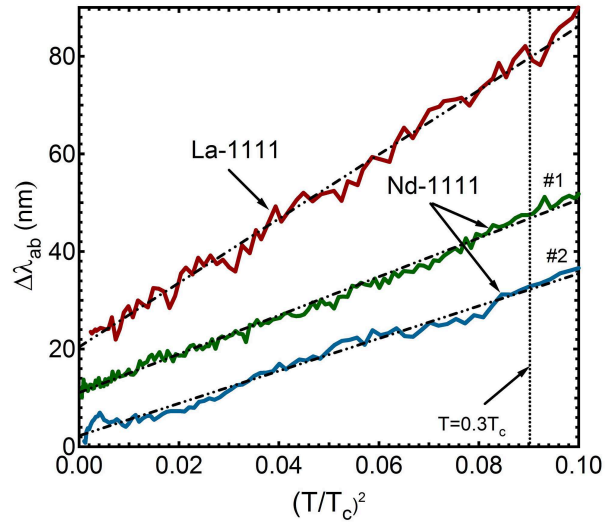}
\caption{(Color online) $\Delta\lambda_L$ vs $(T/T_{c})^2$ after dividing the paramagnetic contribution for two Nd-1111 samples and raw data for a La-1111 crystal (continuous lines). The curves have been shifted vertically for clarity.}
\label{fig:LvsTemp}
\end{figure}

As shown in Fig.~\ref{fig:DivCurr}(b), for both forms the best fits are obtained for $\theta\approx 0.2$~K, which suggests the possibility of antiferromagnetic order of the Nd moments. A fit to the s-wave BCS form Eq.~(\ref{Eq: BCS_Mu}) with $\Delta_{0}$ and the Curie-Weiss constant $C$ as the free parameters yields $\Delta_{0}$=(0.8$\pm$0.2)T$_c$ and $C$=0.085$\pm$0.03. Through many iterations we find that the best convergence to Eq.~(\ref{Eq: Pow_Mu}) is obtained for $n \approx 2$. This power not only reproduces the data accurately, but it gives a narrow range of the variation for $C=0.18\pm 0.02$, which is close to the theoretical $C=0.25$ estimated for the free Nd$^{3+}$ moment of $3.62 \mu_B$. The inset of Fig.~\ref{fig:DivCurr}(b) shows the London penetration depth $\lambda_{L}(T)$ obtained after dividing by $\sqrt{\mu(T)}$ for two sets of values ($C$, $\theta$): (0.085, 0.2) from the best fit to Eq.~(\ref{Eq: BCS_Mu}) and (0.18, 0.2) from the best fit to Eq.~(\ref{Eq: Pow_Mu}). We notice that an assumption of exponential $\lambda_{L}(T)$ does not remove the low temperature upturn. $\lambda(T)$ obtained from the fit to Eq.~(\ref{Eq: Pow_Mu}) does not saturate exponentially for $T\leq 6K$, contrary to earlier reports that analyzed the \textit{total} measured penetration depth ~\cite{Malone08, Hashimoto09, Martin08}. As shown in  Fig.~\ref{fig:LvsTemp}, $\lambda_{L}(T)$ is well described by a quadratic temperature dependence down to 0.02 $T_c$ for all studied R-1111 crystals. This remarkable finding together with the results from Ref.~\cite{Gordon08, Martin09}, suggest that the nearly quadratic temperature dependence of the London penetration depth is a universal characteristic of the iron-pnictide superconductors, not only of the FeAs-122 family. This finding offers compelling evidence for unconventional superconductivity in RFeAsO$_{0.9}$F$_{0.1}$.

\begin{figure}[tb]
\includegraphics[width=8.5cm]{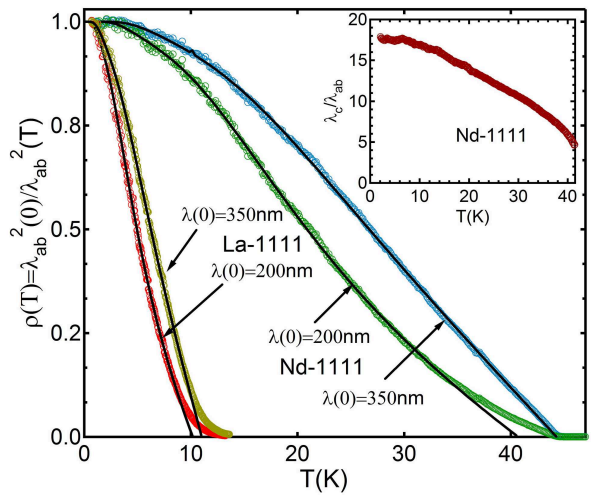}
\caption{(Color online) The superfluid density, $\rho(T)$, for R-1111 crystals calculated with the two values of 200 and 350 nm for $\lambda_L(0)=$ (symbols). Solid lines are fits to the two-gap model described in the text. Inset: temperature dependent anisotropy of the penetration depth.}
\label{fig:RhovsTemp}
\end{figure}

Whereas the pairing mechanism in FeAs superconductors remains an open question, we show below that the penetration depth anisotropy and superfluid density are consistent with the existence of two gaps in R-1111.
To calculate the anisotropy $\gamma_{\lambda}(T)=\lambda_{c}/\lambda_{ab}$ (both quantities are now London penetration depths after accounting for the paramagnetic contribution), we need to know the absolute values of  $\lambda_{ab}(T)$ and $\lambda_{c}(T)$. For the in-plane penetration depth, $\mu$SR experiments give $\lambda_{ab}(0) =200$~nm ~\cite{Khasanov08}. To obtain the absolute value for $\lambda_{c}(T)$, we use the anisotropy of $H_{c2}(T)$ near $T_c$ as described in Ref.~\cite{Tanatar2009}. From our TDR measurements \cite{Martin08} as well as from from resistivity \cite{Jia08} and specific heat \cite{Welp} measurements, $\gamma_{\lambda}=\gamma_{\xi}(T\leq T_{c}) \approx 4-5$ at $T_c$. The inset of Fig.~\ref{fig:RhovsTemp} shows $\gamma_{\lambda}(T)$ calculated assuming $\gamma_{\lambda}(T)=4.5$. Regardless of the initial value chosen, $\gamma_{\lambda}(T)$ increases upon cooling, reaching $\gamma_{\lambda}(0) \approx 17\pm 3$ at $T=0$. Such behavior of the anisotropy was first reported in Sm- and Nd-1111 and interpreted in terms of multiband superconductivity in Ref.~\cite{torque}. A similar temperature dependence was also observed in the FeAs-122 superconductors~\cite{Prozorov09}. A possible explanation for the increase of $\gamma_{\lambda} (T)$ upon cooling is the existence of two superconducting gaps with different magnitudes on the FS with different anisotropies~\cite{Kogan02,torque,Martin09}.

While the low-temperature behavior of $\lambda_L(T)$ is clearly non-exponential, at intermediate temperatures the response is dominated by the quasiparticles thermally excited over the gaps. One can try a simple s-wave model to probe possible two-gap superconductivity, prompted by the positive curvature of $\rho(T)$, similar to MgB$_2$ \cite{Fletcher05}. Figure~\ref{fig:RhovsTemp} shows the in-plane superfluid density, $\rho=\lambda^2(0)/\lambda^2(T)$. Due to uncertainty in $\lambda(0)$ we have used the two values of 200 nm and 350 nm that agree with the literature for the 1111 pnictides \cite{Khasanov08,Luetkens350nm}. To fit the data, we have used the model, where $\rho(T)=\varepsilon\rho_{1}\left( \Delta_{1}\right)  +\left(1-\varepsilon\right)  \rho_{2}\left(  \Delta_{2}\right)$ with two clean s-wave gaps $\Delta_{1}(T)$ and $\Delta_{2}(T)$ along with relative densities of states, $\varepsilon$ and $(1-\varepsilon)$. Successful fits were obtained in all cases with the following values of ($\Delta_{1}/k_B$, $\Delta_{2}/k_B$ and $\varepsilon$): La-1111 (11.29 K, 4.16 K,0.83) for $\lambda(0)=200$ nm and (14.98 K, 5.24 K, 0.82) for $\lambda(0)=350$ nm. For Nd-1111, we obtain (47.25 K, 14.71 K, 0.86) and (64.15 K, 21.61 K, 0.85). In all samples, the ratio between the gaps is $\Delta_{1}/\Delta_{2}\approx 3$, with the large gap at the FS sheet with 80$\%$ of the total density of states. This result is similar to previous reports on Sm-1111 \cite{Malone08} and Pr-1111~\cite{Matano08, Hashimoto09,okazaki09}. This analysis poses two interesting questions. First, for all cases, we obtain $\Delta_{1}/k_{B}T_c$ and $\Delta_{2}/k_{B}T_c$ smaller than the weak-coupling value of 1.76. Second, for the experimentally determined value of $\lambda_{ab}(0)$=200 nm, the superfluid density shows a pronounced positive curvature at intermediate temperatures and a strong suppression near $T_{c}$. In the vicinity of $T_{c}$, the superfluid density should be linear with temperature, $\rho(T)=\eta\left(1-T/T_{c}\right)$. For a clean $s$-wave superconductor $\eta=2$, whereas for RFeAsO$_{0.9}$F$_{0.1}$ we obtain $\eta\approx 1$. This reduced slope agrees qualitatively with the theoretical calculations for the case of two $s$-wave gaps~\cite{Nicol05} and/or extended $s^{+}$ symmetry~\cite{Vorontsov09} when a large interband scattering rate $\tau_{12}$ is considered: $\eta\propto 1/\tau_{12}$, therefore a higher $\tau_{12}$ will lower the slope of $\rho(T)$ at $T_{c}$.

In conclusion, we find that the in-plane London penetration depth in single crystals of RFeAsO$_{0.9}$F$_{0.1}$ (R=La,Nd) follows a power-law temperature dependence, $\Delta \lambda_{ab}(T)\propto T^2$ for $T<T_c/3$. The penetration depth anisotropy $\gamma_{\lambda}$ increases upon cooling and the analysis of the superfluid density suggests two-gap superconductivity. Given our results on 122 family \cite{Gordon08,Martin09}, we conclude that iron-based pnictides are unconventional two-gap superconductors.

We thank D.~S.~Robinson for help with the x-ray experiment, A.~Carrington, A.~V.~Chubukov, R.~W.~Giannetta, I.~I.~Mazin, G.~D.~Samolyuk, J.~Schmalian and A.~B.~Vorontsov for discussions. Work at the Ames Laboratory was supported by the Department of Energy-Basic Energy Sciences under Contract No. DE-AC02-07CH11358. M.A.T. acknowledges continuing cross-appointment with the Institute of Surface Chemistry, NAS Ukraine. R. P. acknowledges support from the Alfred P. Sloan Foundation.

\end{document}